# A simple method to produce almost perfect graphene on highly oriented pyrolytic graphite

M. J. Webb [a], P. Palmgren [b,*], P. Pal [b], O. Karis [b], and H. Grennberg [a]

[a] Dept. of Biochemistry and Organic Chemistry, Uppsala University, P.O. Box 576, 751 23 Uppsala, Sweden

[b] Dept. of Physics and Astronomy, Uppsala University, P.O. Box 516, 751 21 Uppsala, Sweden

**Abstract**

A simple and effective stepwise-method has been developed to remove defects from the top graphene layers of highly orientated pyrolytic graphite. Using a combination of ozone exposure and moderately high temperature we have shown that a defect-rich graphite surface can be modified to generate a graphene-like surface containing a negligible amount of oxygen, hydrogen and $sp^3$ carbon. We report definitive x-ray photoelectron and x-ray absorption spectroscopy analysis after each stage of the process, suggest a mechanism by which the modification occurs and propose it as a route towards the preparation or manipulation of pristine graphene samples.

## 1  Introduction

Graphite is, in the simplest sense, a bulk material consisting of many graphene layers and, as such, an exposed surface of graphite is a model graphene surface on a graphite substrate.

Since Geim[1] and co-workers isolated the first example of graphene in 2004 the race has been on to determine how best to prepare, and make use of, this two-dimensional sheet of carbon

---

[*] Corresponding author: Fax: (+)46-018-471 20 00.
 E-mail address: pal.palmgren@physics.uu.se (P. Palmgren).

whose extraordinary thermal, mechanical and electrical attributes have attracted enormous interest.[1-8]

Research in the field of graphene and pseudo-graphene preparation is vast and recent reviews provide a summary of the enormous volume of work reported in this research area.[9-11] Our own investigations in recent years have focussed on the modification of carbon based materials – nanotubes, $C_{60}$ fullerenes, graphite foil and highly orientated pyrolytic graphite (HOPG) – by use of various solution state and mechanical methods.[12, 13] In an effort to further improve and expand the scope of established graphite modification procedures we are seeking methods that are more selective, efficient and less hazardous than present protocols.[14-18] Reactions where the reagent(s) as well as the by-products are gases fulfils some of our criteria. In the present paper, we give an account of our findings from reacting HOPG with ozone.

Ozone gas is easily generated by ultraviolet excitation of molecular oxygen and the molecule is described as the average of two resonance forms (Figure 1). This molecule is a powerful oxidising agent which can react directly as $O_3$.

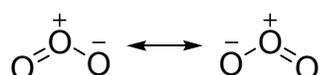

Figure 1. Bonding arrangement in ozone.

Ozone is used prevalently in water disinfection and purification processes where it will oxidise many nitrogen, sulphur and carbon containing compounds.[19, 20] Most importantly though, ozone will react with unsaturated carbon double bonds such as those found in graphene *via* a selective process known as ozonolysis. The key steps, in what is known as the Criegee[21] mechanism, have been confirmed by Berger *et al*.[22]

The benefits of maximising an oxidative process based on ozone are great. The process potentially generates only gaseous waste products, $O_2$, CO and $CO_2$ making experimental work up straight forward. The experiment can be conducted *in situ* thus avoiding troublesome transfer and ambient exposure issues. A further advantage of ozone is selectivity as the gas focuses primarily on exposed surfaces and therefore interferes significantly less with the bulk material.

Previous experimental work incorporating ozone treated graphene or HOPG has been reported to allow uniform atomic layer deposition (ALD) of metal oxides.[23-26] More specific analysis of ozone-treated HOPG has employed scanning tunnelling microscopy to obtain information about the selective chemical etching of the graphite surface by ozone.[27-29] Raman spectroscopy of ozone-exposed graphene samples identified an increase in the *p*-doping levels of the material, without the introduction of significant disorder into the carbon framework.[30] Ozone adsorption on graphene has been shown computationally to be reversible for relatively modest doses.[26] In addition, as model compounds for graphene, the ozone oxidation of polycyclic aromatic hydrocarbons have been investigated *in silico*.[31] However, a thorough understanding of the chemical composition and characterisation of a graphene or a HOPG surface post-ozone treatment is still lacking. Using a HOPG surface as a model graphene system, our work highlights both the initial state of the ozone-graphite surface and the subsequent condition of the samples post annealing at temperatures up to 530 K.

## 2    Experimental

For our graphite starting material we used individual samples (5 mm x 5 mm x 1 mm) of HOPG whose top and bottom most layers were removed by mechanical exfoliation using the scotch tape method.[1] The graphite samples were then sealed in a vial and subjected to a drying process at 333 K under low vacuum (~1 Torr) for 3 hours. The samples were subjected to two ozone exposure times at room temperature, 10 and 60 minutes, before being flushed with and stored under argon, a process considered to quench any further significant oxidative reaction. Finally, the sample vials were again dried for 3 h (1 Torr, 333 K) and stored under argon. These measures were taken to minimise the exposure of our chemically treated samples to the ambient, but nevertheless, some unintentional reactions (e.g. with water) cannot be excluded although high purity gases were used. Ozone was generated using pure $O_2$ and a Fischer 500 UV Ozone generator with flow rates of 40 L/h containing ~ 4 % $O_3$. This equated to an ozone exposure of approximately 2 g/h.

The electron spectroscopy experiments were performed at beam line D1011 at MAX Lab Sweden. This is a bending magnet based beam line housing a Scienta R4000 electron analyzer for the x-ray photoelectron spectroscopy (XPS) experiments and a MCP detector for partial yield x-ray absorption spectroscopy (XAS).[32] Photoelectron spectra are first normalised to background on the low binding energy side of the peak, and in order to enhance spectral features, spectra are normalized to peak height. Energy level calibration is done by measuring Au 4*f* and the Fermi level of an Au foil in electric contact with the sample. The binding energy scale is adjusted so that the Fermi level corresponds to 0 eV binding energy. XA spectra are recorded with the MCP detector operating at a retarding potential of -100 V to remove the inelastic electrons. XA spectra are normalized to $I_0$ measured on a Si photodiode located in the analysis chamber and the pre-edge is set to zero and the high energy side (at 330 eV) is set to a common value for the geometries used in each preparation. The photon energy is calibrated by measuring Au 4*f* with first and second order light and the photon energy resolution is about 200 meV at 300 eV.

## 3 Results and discussion

### 3.1 XPS Characterisation of HOPG samples

Our primary characterisation techniques were synchrotron based XPS and XAS which provide surface specific information on the chemical species present (XPS), and the local bond arrangement and the atomic structure (XAS) respectively. All XP spectra were recorded at room temperature, at a take-off angle of 30° and the photon energy was chosen such to achieve the maximum surface sensitivity. The low photoelectron escape depth and geometry meant that 70 % of the sampling signal was representative of the top monolayer of the HOPG surface. In order to elucidate the chemical situation and determine which groups were present in our samples a deconvolution of each XP spectrum was made. An asymmetric line profile (Doniach-Šunjić) for the $sp^2$ carbon component and a symmetric convolution of a Lorentzian lifetime and Gaussian instrumental broadened (Voigt) line shape function for the other components were used. An integral (Shirley) background was also subtracted to account for the inelastic photoelectrons. The air exfoliated and untreated HOPG sample (UG) was analysed first to provide a benchmark for our study. The C1$s$ core level XPS spectrum (Figure 2 a) is dominated by a single peak (C2) at 284.4 eV assigned to the $sp^2$ carbon framework (surface and bulk), although a foot on the high binding energy (BE) side of the main line was also observed. These spectral features are dependent on the chemical environment experienced by the carbon atoms present on the surface and, as a result, can be assigned to different carbon containing groups. The deconvolution shows that carbon-hydrogen bonded groups in a $sp^3$ hybridised state (C3 and C3') as well as carbon-oxygen bonded groups (C-O-C, C-OH and O=C-O) are present on the surface. The C3 and C3' components at 284.8 eV and 285.2 eV can be associated with $sp^3$ type carbon atoms.

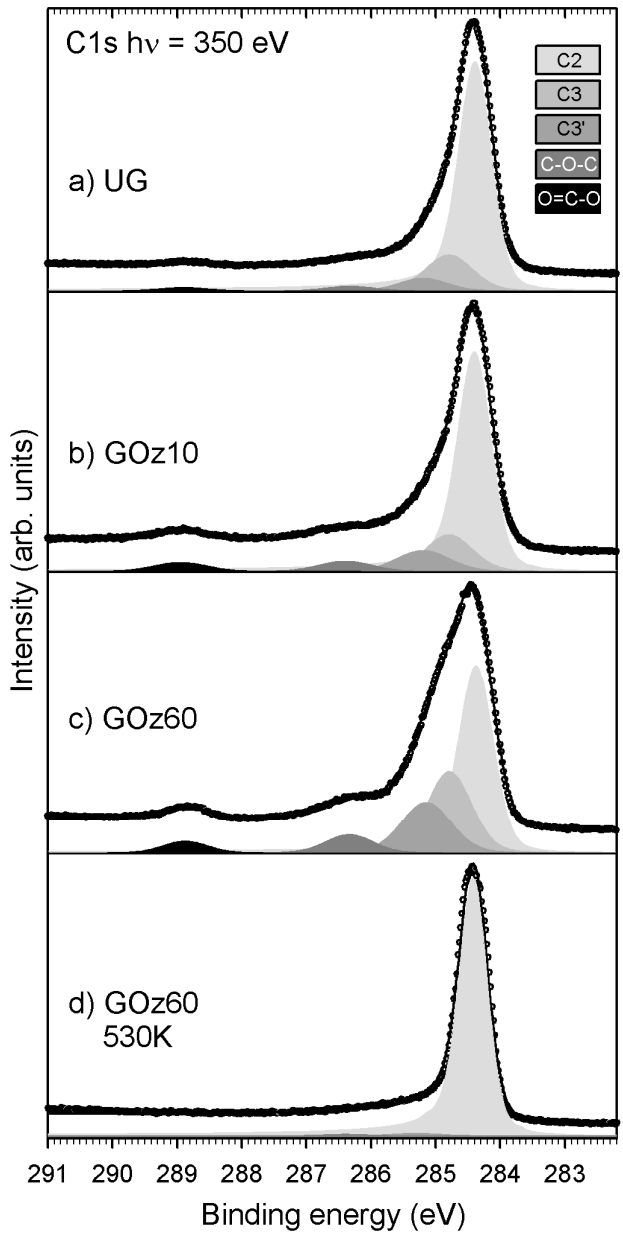

Figure 2. High resolution XPS C1*s* (photon energy 350 eV) spectra of (a) untreated HOPG (UG-298 K), (b) HOPG after 10 min ozone treatment (GOz10), (c) HOPG after 60 min ozone treatment (GOz60) and (d) HOPG after 60 min ozone treatment and annealment at 530 K (GOz60-530 K). The spectra highlight the significant variations in carbon species present on the HOPG surfaces.

In this case, two components were required in order to achieve an accurate fit to the data, but one should imagine several different chemical contributions to these deconvoluted signals. For example, the $sp^3$ carbon of diamond is measured at 284.8 eV[33-35] and the hydrogen termination of this carbon framework does not lead to a large change in binding energy.[33] Hydrogen species adsorbed on HOPG are located at 285.2 eV.[36, 37] Physisorbed hydrocarbons may also contribute to these two features. The C3 and C3' peaks were broader than the C2 component indicative of a slightly more inhomogeneous bonding situation. A more detailed deconvolution of these contributions would be meaningless since there is no appropriate way to model them. The signals at 286.4 eV (C-O-C) and at 288.9 eV (O=C-O) are associated with ether/phenolic groups [38-41] and ester/carboxylic groups,[42, 43] respectively. Finally, the comparative broadness of the C2 peak will be affected by lattice defects and grain boundaries in the graphite material.[44, 45] Annealing the UG sample (not shown) caused a reduction of the non-$sp^2$ C components and an overall increase in the C1$s$ core level intensity. The combination of these observed features indicated a release of physisorbed molecules as well as a degree of surface modification *via* the removal of labile chemisorbed oxygen and hydrogen containing groups (C3, C3', C-O-C and O=C-O). This is well in accordance with expectations as temperatures up to 900 K are routinely used in order to prepare good HOPG surfaces.[36, 46] The oxygen content, as monitored by the O1$s$ signal, provided the most significant changes pre and post heat treatment as a significant reduction in oxygen concentration was observed after annealing. The remaining levels of impurities were around the detection limit (See supplementary information).

## 3.2  Ozone exposed HOPG

The HOPG samples exposed to ozone for 10 and 60 minutes, GOz10 and GOz60, recorded a gradual increase in the C1*s* core level signals in the XP spectra (Figure 2, b and c) indicative of the increased presence of oxidised and $sp^3$ carbon species. The following analysis makes use of the presence of water molecules as a source of oxygen and hydrogen, even though measures were taken to minimize exposure of the samples to the ambient by storage in Ar. Deconvolution of the C1*s* signals identified five distinct components in each case. The largest contribution in both samples came from the $sp^2$ C framework (C2 - 284.4 eV) on the surface and from the bulk of the material (C=C and benzene-like CH terminating the framework). Signals C3 and C3' were characterised as combinations of $sp^3$ C-C and $sp^3$ C-H containing groups whilst two carbon-oxygen environments were identified at 286.4 and 288.9 eV as ether/phenolic components (C-O-C) and ester/carboxylic groups (O=C-O) respectively. Literature reports of IR spectroscopy on ozone exposed graphite powder supports the existence of these carbon-oxygen functionalities.[47] Of notable absence from the XP spectra were ketone groups (~287.5 eV) and carbonate groups (~290.4 eV).[40, 41, 48]

The XP spectra in Figure 2 were normalised to allow the spectral differences between the UG, GOz10 and GOz60 to be compared. Significantly, the C3 and C3' components approximately doubled in concentration after 60 minutes of ozone exposure. Table 1 highlights the various $sp^3$ carbon species present in the GOz10 and GOz60 C1*s* XP spectra and their evolution relating to ozone exposure. The reduction in the percentage of $sp^2$ C present was interpreted as an indication of the progress of the ozone reaction with the surface $sp^2$ C framework.

The degree of oxidation, however, appeared limited and independent of our ozone exposure times (10 and 60 minutes). For both samples the combined intensity of the signals for ether/phenolic (286.4 eV) and carboxylic/ester (288.9 eV) functionality types only reached levels of 8-9 % of the total C1$s$ intensity. In addition, the O1$s$ spectra (see Supplementary Information) identified similar concentrations of oxygen present in both the GOz10 and the GOz60 samples. This result indicated that a surface saturation of oxygen species (physi- and chemisorbed) was observed for both samples. This behaviour would be in agreement with the ALD studies on graphene utilising ozone pre-treatment.[23-26]

Under the experimental conditions chemical reactions involving $sp^2$ carbon in the graphene basal plane would be difficult because of the hindered geometry, however, at sites where greater access is afforded the ozone molecules (e.g. plane edges), oxygen will be covalently incorporated (chemisorbed) *via* the ozonolysis mechanism. During the reaction these 'sites' will possess reactive intermediates such as trioxolanes. The continued influx of ozone could then propagate reactions at these sites breaking up the $sp^2$ C framework further. During this process we believe the reaction generates waste products in addition to the gaseous $O_2$, CO and $CO_2$ expected.[49, 50] In much the same way as fire burns a sheet of paper leaving behind ash, the ozonolysis reaction progresses through the $sp^2$ C lattice generating molecular fragments containing carbon, oxygen and hydrogen which remain on the HOPG surface below. These fragments, mainly $sp^3$ carbon based, are responsible for the increased levels of non-$sp^2$ C species in the C1$s$ XP spectra (C3, C3', C-O-C and O=C-O). The evolution of the signals related to the different species on the surface is summarised in Table 1.

**Table 1.** Relative areas in percent of the deconvoluted components in the C 1s peaks and the binding energies (eV) of the fitted peaks.

| Sample | C2[a] $sp^2$ C=C 284.4 | C3 $sp^3$ C-C and C-H 284.8 | C3' 285.2 | C-O-C Ether/ phenolic 286.4 | O=C-O Ester/ carboxylic 288.9 |
|---|---|---|---|---|---|
| UG | 78.5 | 12.2 | 6.3 | 1.8 | 1.2 |
| GOz10 | 70.2 | 13.4 | 8.5 | 4.2 | 3.7 |
| GOz60 | 50.8 | 23.7 | 16.5 | 5.6 | 3.4 |
| GOz60 530K | 97.8 | - | 1.4 | 0.8 | - |

FWHM ranges 0.46-0.57 eV for C2, and 0.76-0.91 eV for C3, C3', C-O-C & O=C-O. The Lorentzian width is kept at 0.12 eV.

[a] Fitted with an asymmetry parameter of 0.06

Once the flow of ozone is stopped, the chemical attack on the edges of the $sp^2$ carbon network also ceases, leaving behind regions of defect carbon (molecular fragments) and areas, or pits, where the top graphite layer has been removed.[27] At the edges of the $sp^2$ network will be found structural disorder ($sp^3$ C) where the plane of carbon, and any carbon fragments, will be terminated by the various hydrogen and oxygen containing functional groups highlighted in the C1s XP spectrum (Figure 2 and Table 1). The contribution of surface bound water (introduced during the transfer of the sample to the XPS instrument) has been considered in the analysis and, is thought to be the one of the sources of hydrogen necessary for the significant CH related signal observed in the C1s XP spectrum. We believe a second source of hydrogen to be $H_2$ as this is the only element present in the XPS sample chamber at an appreciable concentration under these extremely low pressure conditions.

### 3.3 Thermally treated ozone exposed HOPG

Annealing of the ozone treated samples affected a significant change in the surface composition. Figure 2 (d) shows the C1$s$ XP spectrum after annealing the GOz60 sample to 530 K. This heat treatment caused almost all of the non-$sp^2$ C functional groups to be removed. Although a small percent of the original intensity remained for the C-O-C/C-OH (0.8 %) and $sp^3$ C3' (1.4 %) peaks, these chemical states were only associated with the edges of the $sp^2$ framework, in the form of ethers, phenolic groups and hydrogen bound to carbon in a $sp^3$ hybridised environment. The observed C1$s$ peak width after ozone exposure and annealing amounted to 0.46 eV, comparably smaller than the 0.54 eV FWHM found for the non-ozone treated sample annealed to similar temperatures, see supplementary information. As a result, the combined treatment leads to increased chemical homogeneity at the surface that otherwise would require significantly higher annealing temperatures to achieve.[36, 46] Further to this, the shake-up [51, 52] at about 291 eV becomes stronger which also points towards a well ordered surface of $sp^2$ character.

In addition to these measurements XAS experiments were also carried out. By changing the incident angle of the synchrotron light E-vector to the surface plane, geometric information can be obtained. With an angle of 25° corresponding predominantly to promoting a C1$s$ electron into unoccupied states of local π symmetry, and an angle of 90° probing states of local σ symmetry, significant information can be obtained about the local bond arrangement. Graphite, with its highly ordered flat lying planes of graphene, is an excellent system to study by XAS. The XA spectrum for graphite is dominated by two diagnostic $sp^2$ C framework features; the π$^*$ resonance at 285.5 eV and the σ$^*$ resonance at 291.5 eV.[46]

Our measurements were made in partial electron yield to maximise the surface sensitivity so as to achieve a good correlation with the XPS experiments. The XA spectra recorded of the as-prepared HOPG sample (Figure 3), highlighted the presence of existing irregularities in the surface structure, identified as the signal between the $\pi^*$ (285.5 eV) and $\sigma^*$ (291.5 eV) peaks, and attributed to surface bound carbon contaminants such as singly bound oxygen (C-O) at 287.0 eV[53, 54], hydrogen (C-H) at 287.6 eV[36, 37] and oxygen (O=C-O) found at 288.5 eV[54, 55] as well as existing irregular surface structures. This portion of the spectrum has similarities with XA spectra recorded from diamond[56], a system in which no $sp^2$ bound carbon exists, hence the upwards slope between the $\pi^*$ and $\sigma^*$ peaks can be associated with $sp^3$ bound carbon.

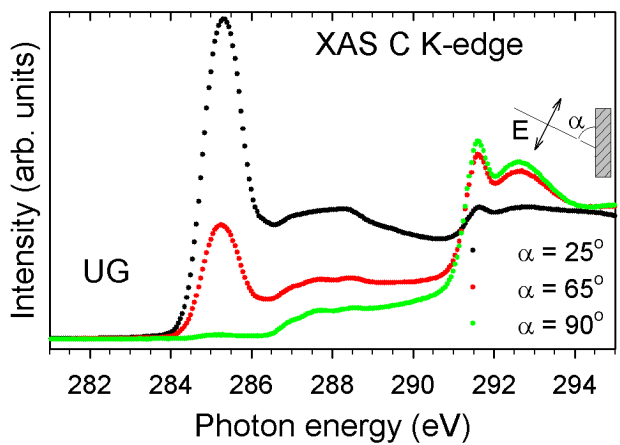

Figure 3. XA spectra of untreated HOPG (UG). The $\pi^*$ transition can be clearly seen at 285.5 eV (25°) and the $\sigma^*$ transition at 291.5 eV (90°).

The XA spectra recorded for the HOPG sample exposed to 60 minutes of ozone (GOz60) can be seen in Figure 4 (a) where ozone exposure lead to an increase in intensity in the $sp^3$ related region between the $\pi^*$ and $\sigma^*$ peaks. The oxygen related state at 288.5 eV was clearly seen and the slope from $sp^3$ carbon became steeper, in agreement with the observations in the C1$s$ spectra (Figure 2).

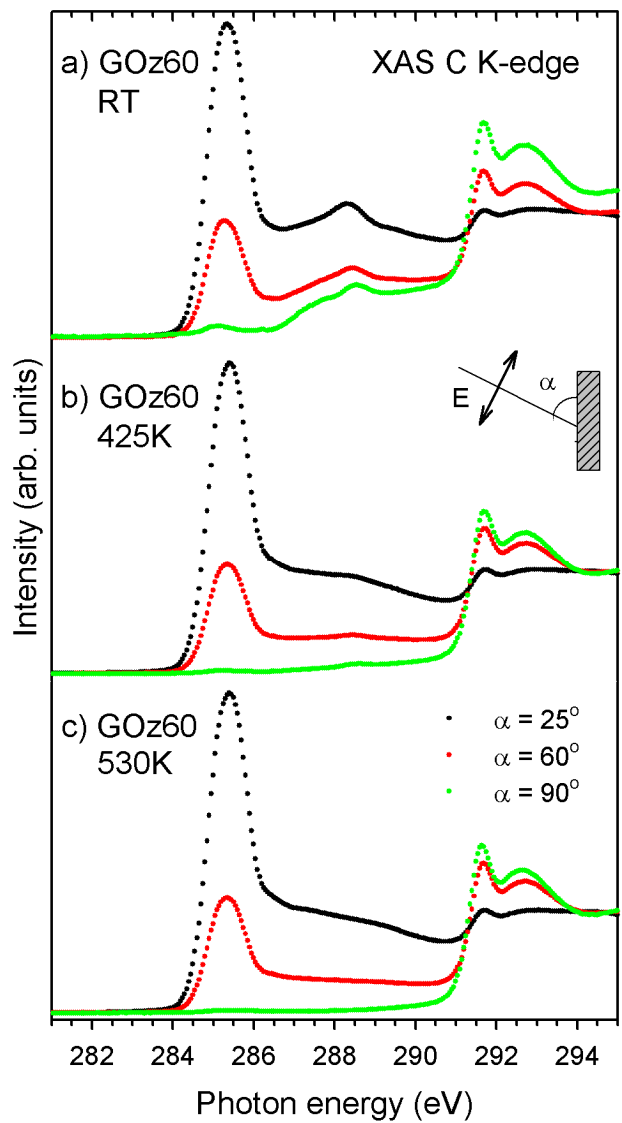

Figure 4. XA spectra of (a) HOPG after 60 min ozone exposure (GOz60), (b) after annealment to 425 K (GOz60-425K) and (c) after annealment to 530 K (GOz60-530K).

Figure 4 (b) shows the spectra after the first annealing step of GOz60 at 425 K. The concentration of oxygen and hydrogen containing species, depicted by the signal in the defect region, was significantly reduced. Finally, Figure 4 (c) displays the spectra acquired after heating the sample to 530 K (GOz60-530K). The oxygen and hydrogen carbon-containing components were almost completely removed. The signal related to the $sp^3$ carbon atoms in the XA spectra identified in the original UG sample (Figure 3) as well as in the GOz60 sample completely disappeared. This is not the case for UG annealed to 425 K, in which the $sp^3$ related signal remains (see Supplementary Information). The results of the heat treatment of GOz10 (not shown) were very similar to those obtained for GOz60 (Figure 4).

## 4 Proposed reaction pathway

The observed behaviour was attributed to the ozone treatment attacking overly exposed regions of carbon (e.g. grain boundaries and step edges) and generating $sp^3$ C waste products. The reactions occurred with the upper-most layers of the HOPG before the thermal treatment removed the remaining physisorbed material and the waste products of the ozonolysis reaction. The systematic study of the HOPG ozone-heat treatment process produced a series of high-resolution XP and XA spectra that have allowed us to propose the following.

Figure 5 highlights the proposed selective ozone etching process. The untreated HOPG schematic (a) depicts the top layer of graphene (L1) containing two defects, the second layer of graphite (L2) and the remainder of the HOPG sample as bulk graphite (black). The light grey areas at the edges represent the chemisorbed, predominantly $sp^3$ hybridised, carbon in the form of C-H, ethers and phenolic (C-OH) groups. The $sp^2$ C species (C=C and C-H) present on the surface are shown in the slightly darker grey. Using the same colour scheme, the adjacent expansion shows the proposed edge structure of the top graphene layer, L1. In addition to the $H_2$ present in the sample chamber, water is shown as another source of hydrogen for the reactions.

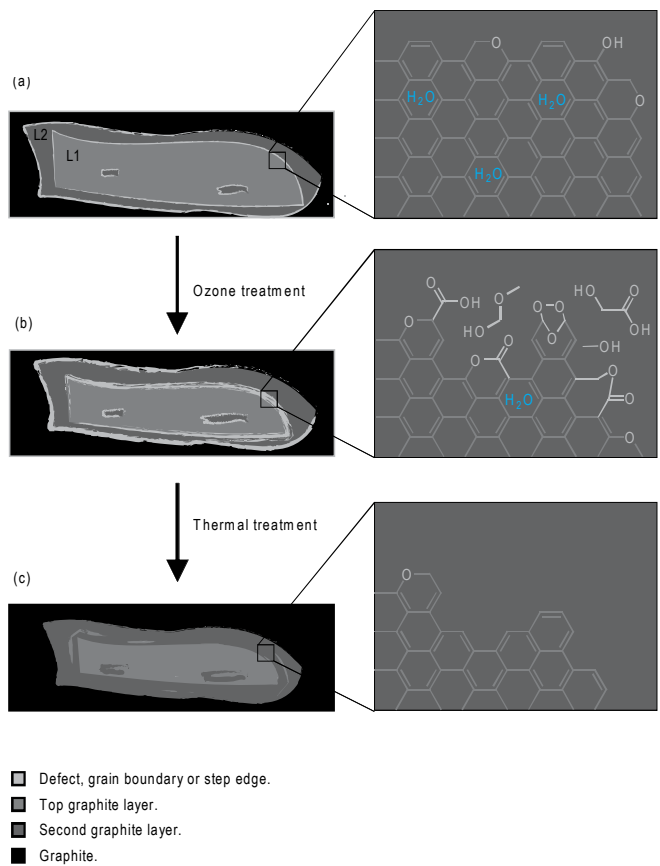

- ☐ Defect, grain boundary or step edge.
- ☐ Top graphite layer.
- ☐ Second graphite layer.
- ■ Graphite.

Figure 5. Our proposed ozone chemical etching of a graphite surface. (a) untreated HOPG surface, (b) surface after ozone treatment and (c) surface after ozone and thermal treatment.

The proposed pathway by which the ozone attacks the graphene mono-layer is depicted in Figure 5 (b). Here it is shown that ozone, *via* an ozonolysis-type mechanism, preferentially reacts with the most exposed regions of carbon e.g. layer edges, grain boundaries or pre-existing defect sites. The possible types of oxygen containing functionality and potential waste products are displayed in the expansion. Also drawn here is the trioxolane species described in Criegee's mechanism.[21] Whilst the sample is in the presence of ozone we believe this etching process to be 'ongoing'. The $sp^2$ carbon network is slowly broken-up leaving behind carbon fragments containing oxygen and hydrogen. Although physisorbed ozone species are likely adsorbed on the graphite basal plane we do not believe, under these relatively mild conditions, that any chemical modification of the bulk $sp^2$ carbon framework occurs. Figure 5 (c) represents the condition of the surface after annealing the sample to moderately high temperatures under vacuum. These experimental conditions remove all the defect carbon (molecular fragments) generated in the ozonolysis reaction as well as a significant proportion of the remaining carbon-oxygen species from the edges of the graphene layers. The resulting top mono-layer consists mainly of a $sp^2$ C lattice whose edge is terminated by a combination of benzene-like $sp^2$ CH with $sp^3$ CH and C-O-C groups as supported by Figure 2 (d). The almost complete absence of spectral features between the $\pi^*$ and $\sigma^*$ resonances in the XA spectrum (Figure 4 c) is evidence that the process removes any labile or physisorbed species from the graphite surface.

## 5 Conclusion

Our experimental investigations and in depth x-ray photoelectron spectroscopy analysis, utilizing HOPG as a model graphene system, have provided information on the composition of a chemically untreated mechanically prepared graphite surface, as well as the condition of that surface after ozone and thermal treatments. Interpretation of the data suggests significant benign modifications had been made to the surface. We found that the HOPG surface post ozone treatment, contained combinations of carboxylic acids and ester groups, ethers, C-OH, and CH groups with carbon in both $sp^2$ and $sp^3$ hybridised forms in agreement with the literature.[47] The presence of ketone groups and carbonates was not supported by our data.

Ozone reacts, *via* an ozonolysis-type mechanism, with pre-existing defect sites, grain boundaries and step-edges of the exposed $sp^2$ C framework of the graphite surface. This discrete chemically driven surface etching process results in the introduction of carbon-oxygen and carbon-hydrogen functionality at these specific surface locations. Continued $O_3$ exposure propagates reactions at these sites, generating waste products as the reaction progresses through the framework and facilitates the removal of some carbon in the form of CO and $CO_2$. Quenching of this 'ozonolysis process' is achieved by flushing the system with argon, which leaves the $sp^2$ framework edges terminated with relatively stable carbon-oxygen and carbon-hydrogen functionality. Subsequent thermal treatment in vacuum then facilitates the removal of all the waste products and all but the lowest concentrations of oxygen. This leaves behind a surface/monolayer consisting of $sp^2$ carbon devoid of significant amounts of unwanted functionality and terminated predominantly by $sp^2$ C-H. This clean and simple process provides a route towards uncovering planes of highly ordered graphene, and establishes a methodology that may well be applicable to removing unwanted species or defects from graphene sheets supported on other substrates.

**Acknowledgements**

The authors acknowledge financial support from the Uppsala University Quality and Renewal program for graphene and the Swedish Science Foundation.

**A simple method to produce almost perfect graphene on highly oriented pyrolytic graphite**

M. J. Webb, P. Palmgren, P. Pal, O. Karis, and H. Grennberg

**Supplementary Information**

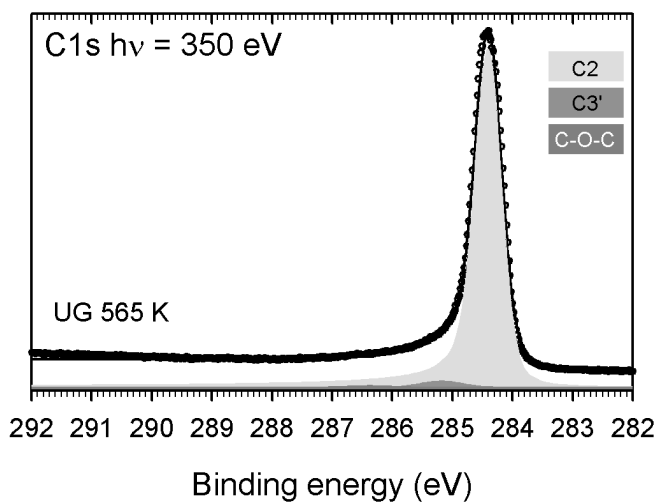

Figure S1: C1$s$ spectrum of UG annealed to 565K; the C3' and C-O-C components remain but are strongly reduced, 2.3% and 0.8% remain respectively. The C3 and O-C=O components are completely gone.

Table S1: Integral areas of the O1$s$ signals for the three samples UG, GOz10 and GOz60.

| Sample | Preparation | Area |
|---|---|---|
| UG | RT | 0.92 |
| | 415 K | 0.33 |
| | 565 K | 0.27 |
| GOz10 | RT | 3.02 |
| | 380 K | 1.25 |
| | 500 K | 0.37 |
| GOz60 | RT | 1.97 |
| | 425 K | 0.64 |
| | 530 K | 0.16 |

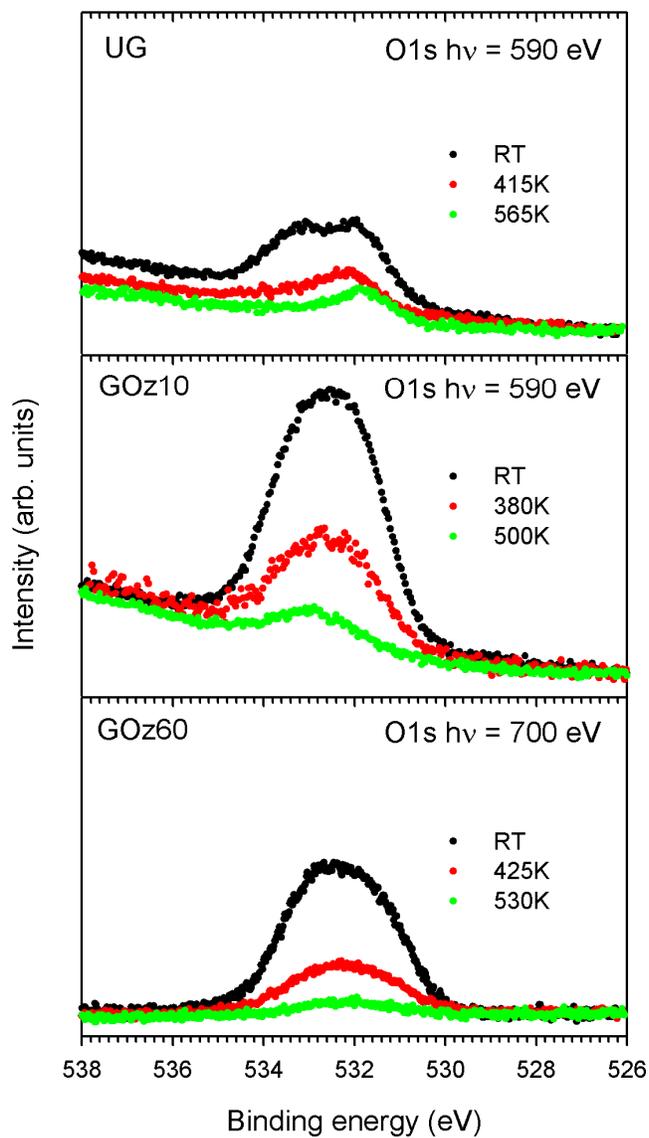

Figure S2: O1*s* spectra from the three samples UG, GOz10 and GOz60, showing the reduction of the oxygen signal with temperature. Note the difference in photon energy between GOz10 and GOz60, leading to a difference in area (see also Table S1) but the oxygen related components in the C1*s* XP spectra are about equal. Exposure to the ambient is roughly equal for the two samples.

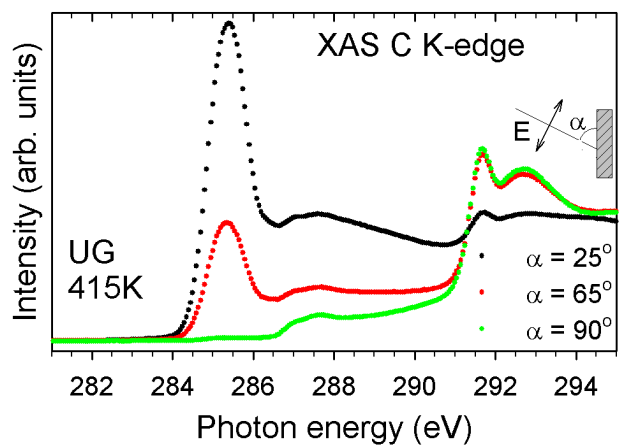

Figure S3: XA spectrum from UG annealed at 415K and the relatively strong contribution from $sp^3$ type carbon between the π* and σ*, this is to be compared to Figure 4b showing the 60 minutes ozone treated sample annealed at 425K.